# Establishing a Learning Model for Correct Hand Hygiene Technique in a NICU

Irén A. Kopcsóné Németh [1,2], Csaba Nádor [3,4], László Szilágyi [5,6,7], Ákos Lehotsky [6,8] and Tamás Haidegger [6,9,10,*]

1. BCE Doctoral School of Business and Management, Corvinus University of Budapest, 1093 Budapest, Hungary; nemeth.iren@hm.gov.hu
2. Medical Centre, Hungarian Defense Forces, 1134 Budapest, Hungary
3. Medical Centre, Hungarian Defense Forces, Site 2, 1068 Budapest, Hungary; nadordr@gmail.com
4. Obstetrics and Gynecology Clinic, Semmelweis University, 1082 Budapest, Hungary
5. Department of Electrical Engineering, Sapientia Hungarian University of Transylvania, 540485 Tîrgu Mureş, Romania; lalo@ms.sapientia.ro
6. HandInScan Zrt., 4025 Debrecen, Hungary; lehotsky@ooi.hu
7. John von Neumann Faculty of Informatics, Óbuda University, 1034 Budapest, Hungary
8. National Institute of Oncology, 1122 Budapest, Hungary
9. University Research and Innovation Center (EKIK), Óbuda University, 1034 Budapest, Hungary
10. Austrian Center for Medical Innovation and Technology (ACMIT), 2700 Wiener Neustadt, Austria

* Correspondence: haidegger@irob.uni-obuda.hu

**Abstract:** The ability of healthcare workers to learn proper hand hygiene has been an understudied area of research. Generally, hand hygiene skills are regarded as a key contributor to reduce critical infections and healthcare-associated infections. In a clinical setup, at a Neonatal Intensive Care Unit (NICU), the outcome of a multi-modal training initiative was recorded, where objective feedback was provided to the staff. It was hypothesized that staff at the NICU are more sensitive towards applying increased patient safety measures. Outcomes were recorded as the ability to cover all hand surfaces with Alcohol-Based Handrub (ABHR), modelled as a time-series of measurements. The learning ability to rub in with 1.5 mL and with 3 mL was also assessed. As a secondary outcome, handrub consumption and infection numbers were recorded. It has been observed that some staff members were able to quickly learn the proper hand hygiene, even with the limited 1.5 mL, while others were not capable of acquiring the technique even with 3 mL. When analyzing the 1.5 mL group, it was deemed an insufficient ABHR amount, while with 3 mL, the critical necessity of skill training to achieve complete coverage was documented. Identifying these individuals helps the infection control staff to better focus their training efforts. The training led to a 157% increase in handrub consumption. The setting of the study did not allow to show a measurable reduction in the number of hospital infections. It has been concluded that the training method chosen by the staff greatly affects the quality of the outcomes.

## 1. Introduction

Hand hygiene (HH) is likely the most efficient tool to fight healthcare-associated infections (HAIs), and also to prevent global pandemics [1]. The survival outcomes among critically ill preterm surgical neonates are very dim and infections are among the leading cause of their mortality [2]. While some articles reported lowered HAI ratios during the ongoing COVID-19 related emergency state [3], in general, infection numbers have grown worryingly, despite the overall increased mask wearing and hand hygiene practices, according to the CDC (https://www.sciencedaily.com/releases/2021/09/210902124943.htm, accessed on 15 June 2022) and International Reports [4,5].

In recent years, the spread of Alcohol-Based Hand Rubs (ABHRs) has shown a lot of benefits, including reduced infection rates, better clinical outcome, while also reducing water usage and the need for infrastructure [6]. In order to perform hand disinfection properly, the complete coverage of the hands' surfaces with an ABHR is essential [7]. Nevertheless, the application of an ABHR requires more skill than HH with soap and water, since there is no flow effect to assist with the cleaning of all areas of the hand, therefore training is critical [8]. Incomplete hand disinfection may severely compromise patient safety, leading to unnecessary medical complication, costs, or fatalities [9]. Ignorance on an incomplete HH technique leads to a patient safety issue, as the most frequently omitted hand areas (thumbs and fingertips) are also high-touch surfaces [10]. During handwashing, application technique is less important from the perspective of efficiency, because of the use of "foaming" and running water, while in the case of hand rubbing, where the hand is not in contact with alcohol for enough time, it would not be disinfected efficiently [11,12].

The typically employed, generic patient safety training has its limitations. Despite the general rising awareness of the importance of SSI prevention, registered data has long shown a rise of HAI again in Hungary [13] (Figure 1), while the data was only made available combined for multi-resistant infections (MR), Clostridium difficile (CDI), and bloodstream infections (BSI). The rising number of incidences once again drew the attention to the importance of effective training. Moreover, there is no recent data available from the National Nosocomial Surveillance System since the COVID-19 outbreak, yet early reports were suggesting that an astonishing 30.6% of COVID-19 patients in the first wave in Hungary (Q1 2020) acquired COVID during their hospitalization (https://444.hu/2020/07/16/minden-negyedik-magyar-beteg-korhazi-fertozeskent-kapta-el-a-koronavirust, accessed on 15 June 2022). Arguably, digital health technologies proved to be very efficient in acting against the coronavirus pandemic [1,3,4,14–16]. Evidence-based medicine is getting well established in various domains of medicine, where accurate outcome measures or digital tools are available for objective data collection [17].

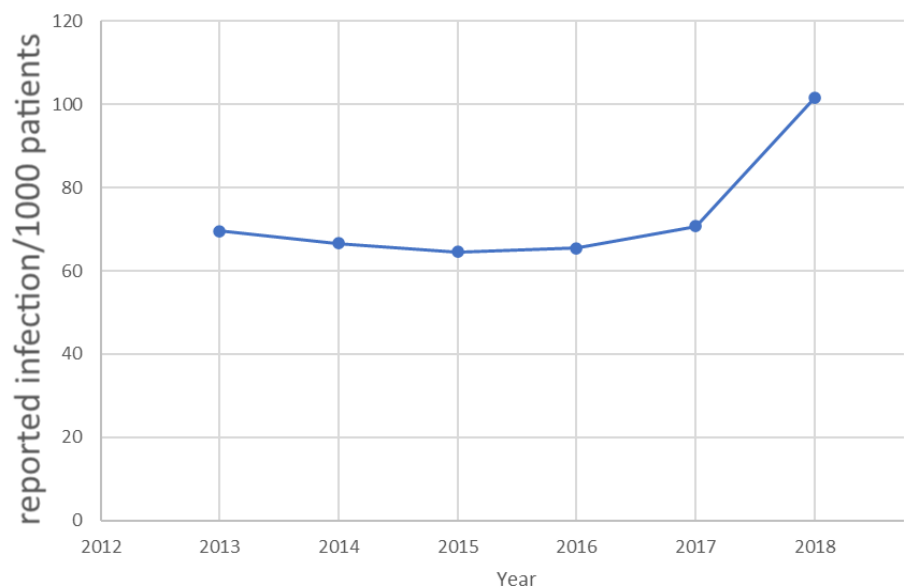

**Figure 1.** The total number of reported infections in Hungary until the Coronavirus outbreak, when the data reporting scheme was fundamentally changed by the authorities. (Source: OEK 2019, [13]).

The Infection Prevention and Control (IPC) staff has been operating an HH improvement project at the 14-bed Neonatal Intensive Care Unit (NICU) of the 1800-bed tertiary Medical Center, Hungarian Defense Forces (Medical Centre) since 2011, based on the World Health Organization (WHO) guidelines [18]. While some countries (e.g., France and parts of China) populate a 7th step for HH (i.e., including the wrists [19]), other publications suggest that it might be disadvantageous from the microbiological efficacy point of view [9]. It was decided to follow the original WHO protocol, which is also the official Hungarian national guideline. The HH training followed a modular approach: besides providing the basic infrastructural conditions (i.e., dispensers, HH stations), the implementation was supported by continuous trainings (for new workers, vocational, and regular trainings) and monitoring (registration of hand disinfectants usage, direct observation compliance audit).

Since 2012, yearly audits have been maintained at the NICU following the NEO-KISS protocol (https://www.nrz-hygiene.de/en/surveillance/hospital-infection-surveillance-system/neo-kiss/, accessed on 15 June 2022). Data collection for quality metrics included the monthly ABHR consumption of all involved departments and the execution of training exercises at various level. In 2016, HH was added to the key training areas, including updated education sessions and the implementation of various related trials [20]. However, the training outcomes have not been systematically evaluated.

The Medical Center's IPC staff has been dedicated to extended HH trainings together with the legally required safety trainings, targeting separately new (incoming) employees (~1000 people per year), organized as monthly 45-min sessions, focusing on the WHO multi-modal strategies and 6-step technique. Numerous training sessions are required to override prior bad imprints, such as incomplete HH technique (which may come from childhood). HH and patient safety trainings are offered 2–3 times a week to existing employees, residents, and visiting clinical staff, leading to a significant time dedication of the IPC staff, therefore optimizing the use of their resources has been a key interest.

Scientific evidence confirms the domain of infection prevention as well as that a single training event cannot change behavior [21]. Recent studies showed that repetition of training for 5–32 times may be sufficient in HH to acquire a certain skill [20,22,23]. In accordance with the concept of deliberate practice in medicine [24], our study focused on the establishment of a learning model in HH, based on the behavior patterns observed in the NICU. It is understood that learning can be most efficient with a feedback loop provided [25,26], therefore the study was focusing on the effectiveness of the learning process.

*Problem Statement and Hypothesis*

During the HH direct observation at the NICU, it was noticed that during caretaking, the routinely done hand rubbing at the bed side was not typically performed following any trained protocol (i.e., the WHO six-step protocol) (Figure 2), thus the fingertips and the thumbs were often left out. Furthermore, the amount and application of the ABHR was very random, with staff pressing the dispenser 1–8 times. This is further complicated by the recent findings by Bansaghi et al. that many clinically installed ABHR dispensers distribute significantly less handrub than their nominal value [27]. As observed, the staff did the rubbing mostly using a common handwashing technique, even though they all had completed a training with the WHO technique.

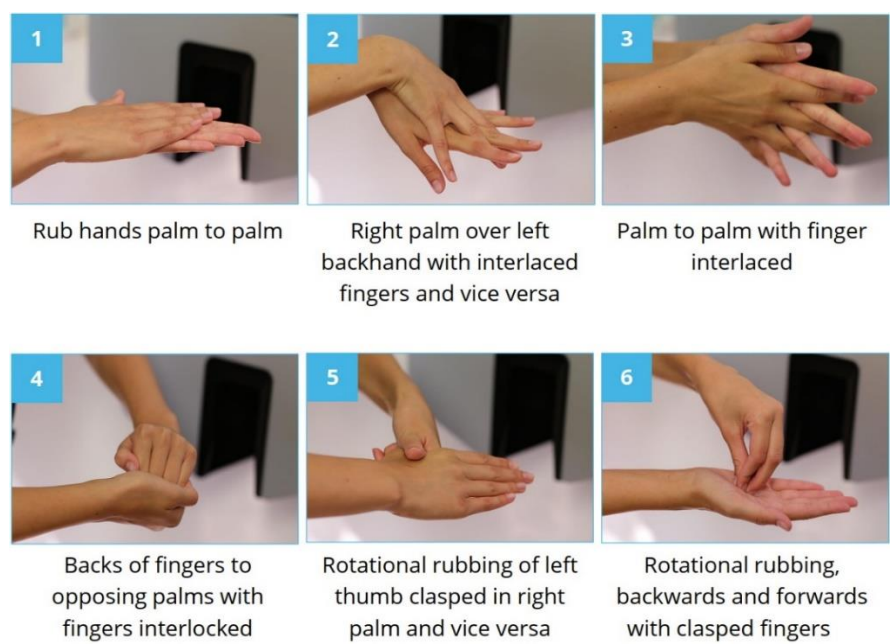

**Figure 2.** The correct WHO 6-step hand disinfection technique. (Source: HandInScan Zrt.).

It became obvious that the traditional training practice needs to be rethought, based on the method called "feedback loop" by psychologists: when more information is provided about something, people will be more likely to follow it [25]. The other practical problem was the over-dosage of the disinfectant. Some workers balanced their technical deficiency with multiple ABHR dosage, which led to very high amounts of used disinfectant, and caused a raise on the financial side of some departments, particularly at the NICU [28].

This study intended to establish basic cost optimization rules for infection control actions, by applying electronic equipment that uses a validated technology [29].

(1) The main institutional level risk is that imperfect HH is dangerous for the patients and the staff; and with ABHR, the correct technique needs to be learned: the goal is to achieve perfect HH, and thus reduce the risk of infection;
(2) Hospitals invest a lot of energy in staff training, but we do not know how effective it is (typically, their shortcomings can be observed easily): the goal is to know and apply effective training methods;
(3) There is a large deviation in ABHR consumption: while 1 mL is a straight precursor for inadequate hygiene, 8 mL is waste—therefore the hospital's goal is to reach an optimal level and establish the range for the staff that guarantees complete HH.

Theoretically, all HCWs shall be able to demonstrate and practice the correct HH technique, but the quantity of necessary ABHR varies from person to person. The applied methodology had a double goal: besides exercising hand rubbing and learning the correct steps of the HH procedure, each participant could establish whether 1.5 mL ABHR solution is sufficient for a perfect HH performance. This was further investigated in a parallel study, primarily focusing on the effect of the pandemic on patient safety methods [30].

## 2. Materials and Methods

### 2.1. Measurements

As a solution for the above stated two problems, with respect to the education of the staff, we started a new training supporting the "feedback loop" method, using the Semmelweis scanner [31].

This study covered the NICU of the Medical Centre and was performed during an 8-week period in 2016. One physician and the chief nurse were designated to supervise the measurements in the ward. The study was initiated with a one-day-long training, during which the participating healthcare workers (HCWs) received Radiofrequency Identification (RFID) cards for anonymous personal identification, completed a survey, and the size of their hands was recorded [32]. HCWs were included anonymously in the study on a voluntary basis, and were identified by the number of their RFID cards. This number connected each participant to their surveys. The survey collected information on sex, age range, work position, years spent in the institution, and the dominant hand. Participants signed an informed consent for inclusion. The contour of each hand was drawn on a template sheet with 1 mm units, which allowed the computation of the hand size (projection area) through computerized image processing. The protocol was presented to the board a priori.

The Semmelweis Scanner (HandInScan Zrt., Debrecen, Hungary), a digital hand hygiene technique assessment tool, was continuously operating in the ward, and the HCWs were invited to test their hand rubbing technique during each shift. Each test started with checking in with their RFID card, followed by performing hand disinfection with an ultraviolet (UV)-labelled ABHR solution Visirub (Bode Chemie, Hamburg) according to the WHO 6-step protocol. The scanner took digital images of both sides of the hands under UV-A light, and produced instant visual feedback regarding the coverage of the disinfectant. The users could see the evaluated images on the screen, and thus had the opportunity to learn how to avoid mistakes [28]. Every participant was allowed to use the device only once a day. The device stored all images together with the RFID information for further offline processing of the individual records. Images were analyzed one by one, and the outcome (percentage of coverage) was recorded.

The dispenser provided the ABHR in doses of 1.5 mL, closest to the average use measured in the wards. The dispenser was equipped with a counter, which was connected to the Semmelweis Scanner. This way, the scanner was able to record ABHR use based on the time, and compute the amount for each measurement.

During the period of the study, HCWs were asked to use only 1.5 mL ABHR each time during the first 3 weeks, and then increase the volume to 3 mL for all further HH events [33].

### 2.2. Statistical Analysis

Statistical analysis was performed using R version 3.1.1 (The R Foundation for Statistical Computing, Vienna, Austria) and MATLAB version 2015a (The MathWorks Inc., Natick, MA USA). Effects were considered significant at $p < 0.05$. Confidence intervals were computed using the Wilson method.

### 2.3. Learning Modelling

The series of HH measurements can also be interpreted as series of state changes. There are two states: adequate (○) and inadequate coverage (×), and thus there are four possible state changes that can be set up as a time series. Markov model analysis was applied, with the assumption that the next state only depends on the current state and not at all on the history of the system. Analyzing the learning abilities of HCWs enables to choose the right educational and teaching method, target future training to the bottom quarter, and thus to optimize IPC resources.

On a monthly base, ABHR consumption is reported per unit, along with patient days, compulsory reported HAI occurrences, staff hours, and other key indicators. The data was partially recorded by the IPC staff, during the study, and also recorded into the Hospital Information System of the Medical Center.

### 3. Results

Thirty-nine HCWs participated in the study. Their survey information is summarized in Table 1. All participants used 1.5 mL ABHR at the first few (3 to 13) measurements, and most of them continued the study with up to 15 measurements using 3 mL ABHR. The measurements of each participant were collected into two chronologically ordered data series. The first series included all measurements of the HCW performed with 1.5 mL and the second with 3 mL ABHR. HH events were declared adequate in case of >95% coverage was achieved, while any other outcome counted as inadequate hand rubbing.

**Table 1.** Statistics of the survey information regarding all participants.

| Classification | | Physicians | Other HCWs |
|---|---|---|---|
| **Gender** | Male | 3 | 0 |
| | Female | 7 | 29 |
| **Dominant hand** | Left | 1 | 3 |
| | Right | 9 | 26 |
| **Age** | <25 | 0 | 2 |
| | 26–35 | 4 | 0 |
| | 36–45 | 4 | 12 |
| | 46–55 | 2 | 14 |
| | >56 | 0 | 1 |
| **Total participants** | | 10 | 29 |

Figure 3 presents the average rate of adequate coverage of all participants, separately identified for each measurement using 1.5 mL and 3 mL ABHR. Figure 3 shows the first measurement series (up to 10 measurements) of those participants, who have performed at least 4 measurements with 1.5 mL ABHR. The first series shows a rising value of the rate during the first 7 measurements, which stabilizes thereafter, indicating that up to 60% of the HCWs can learn the correct hand rubbing technique using 1.5 mL ABHR. The second series suggests that 3 mL of ABHR is enough for any participant to achieve perfect coverage. However, mistakes occur sporadically, even after a long training session.

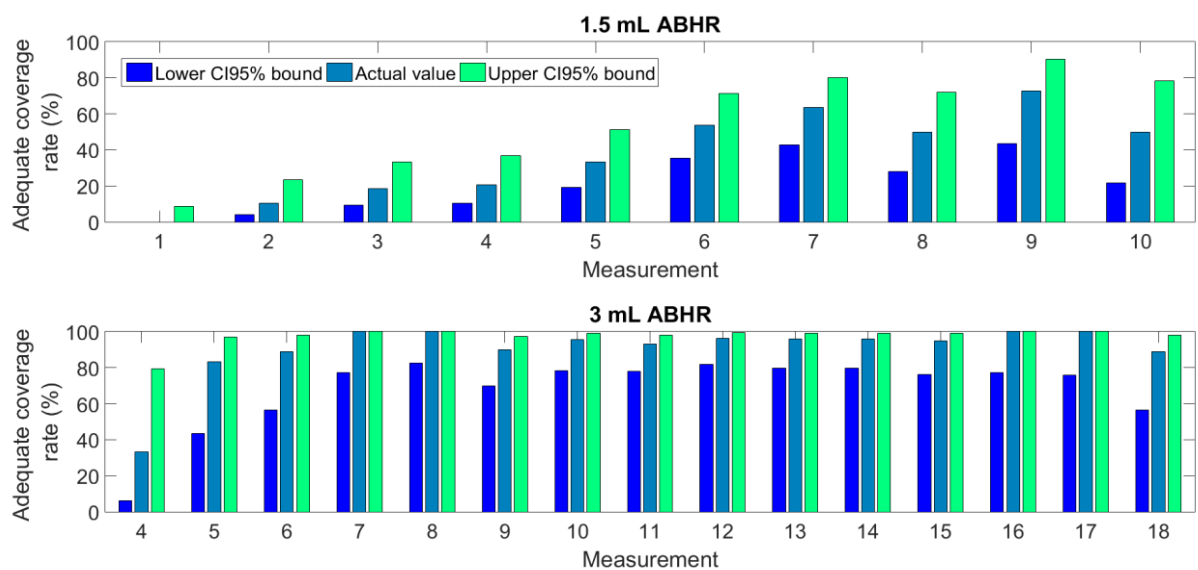

**Figure 3.** The rate of adequate coverage (with CI 95% values) of all participants at various measurements using 1.5 mL and 3 mL ABHR.

The first series of measurements using only 1.5 mL of ABHR solution revealed that the rate of correct HH performances considerably increases starting from the third event and stabilizes after the eighth measurement. This confirms the similar previous findings by Ref. [29], a study that evaluated the role of visual feedback without monitoring the amount of ABHR taken.

The series of measurements exhibited in Figure 4 separately for each HCW can also be interpreted as series of state changes. Figure 5 shows the statistics of state changes, following measurements occurring after adequate and inadequate coverage separately, indicating a learning process. The rate of adequate coverage obtained after an inadequate one at the previous measurement gradually rises above 50% during the first six performances. At any stage of the study, the chances for an adequate coverage at the next measurement were significantly better after having performed adequate coverage at the previous measurement (Figure 6). Markov model analysis, which supposes that the next state only depends on the current state and not at all on the history of the system, estimates a 43% rate for adequate coverage in steady state. We consider this a pessimistic estimation or an underestimated rate, because in contrary to Markov systems, HCWs can learn from every previous mistake.

| Those who were able to learn adequate hand rubbing | Those who were unable to learn adequate hand rubbing |
|---|---|
| Participant | Measurement 1–10 | Participant | Measurement 1–10 |
| 1 | ×↗○→○→○↘×↗○↘×↗○→○→○ | 19 | ×→×→×→×→×↗○↘×→×→×↗○ |
| 2 | ×→×→×→×→×↗○→○↘×↗○↘× | 20 | ×→×→×→×→×↗○↘×→×→×→× |
| 3 | ×→×→×→×→×→×↗○↘×↗○→○ | 21 | ×→×↗○↘×→×→×→×→×→×→× |
| 4 | ×→×→×→×→×→×↗○→○→○↘× | 22 | ×→×→×→×↗○↘×↗○↘×↗○ |
| 5 | ×→×→×→×→×→×→×→×↗○→○ | 23 | ×→×→×→×→×→×↗○↘× |
| 6 | ×↗○↘×↗○→○→○→○→○→○ | 24 | ×→×→×→×→×→×↗○ |
| 7 | ×→×→×↗○→○→○→○→○→○ | 25 | ×→×→×→×→×→×→× |
| 8 | ×→×→×→×→×→×↗○→○ | 26 | ×→×→×→×→×→× |
| 9 | ×→×↗○→○↘×→×↗○→○ | 27 | ×→×→×→×→×↗○ |
| 10 | ×→×→×→×→×↗○→○→○ | 28 | ×→×→×→×↗○ |
| 11 | ×→×→×→×↗○→○→○ | 29 | ×↗○↘×→×→× |
| 12 | ×→×→×→×→×↗○→○ | 30 | ×→×↗○↘×→× |
| 13 | ×→×↗○→○→○→○→○ | 31 | ×→×→×→×→× |
| 14 | ×→×→×→×↗○→○ | 32 | ×→×→×→× |
| 15 | ×→×→×↗○→○→○ | 33 | ×→×→×→× |
| 16 | ×→×→×→×↗○→○ | 34 | ×→×→×→× |
| 17 | ×→×↗○↘×↗○→○ | | |
| 18 | ×→×↗○→○ | | |

**Figure 4.** Measurement time series of individual HCWs performed with 1.5 mL ABHR, represented with a Markov model. Symbols ○ and × stand for adequate and inadequate coverage, respectively. Arrows highlight the changes between the outcomes of consecutive measurements.

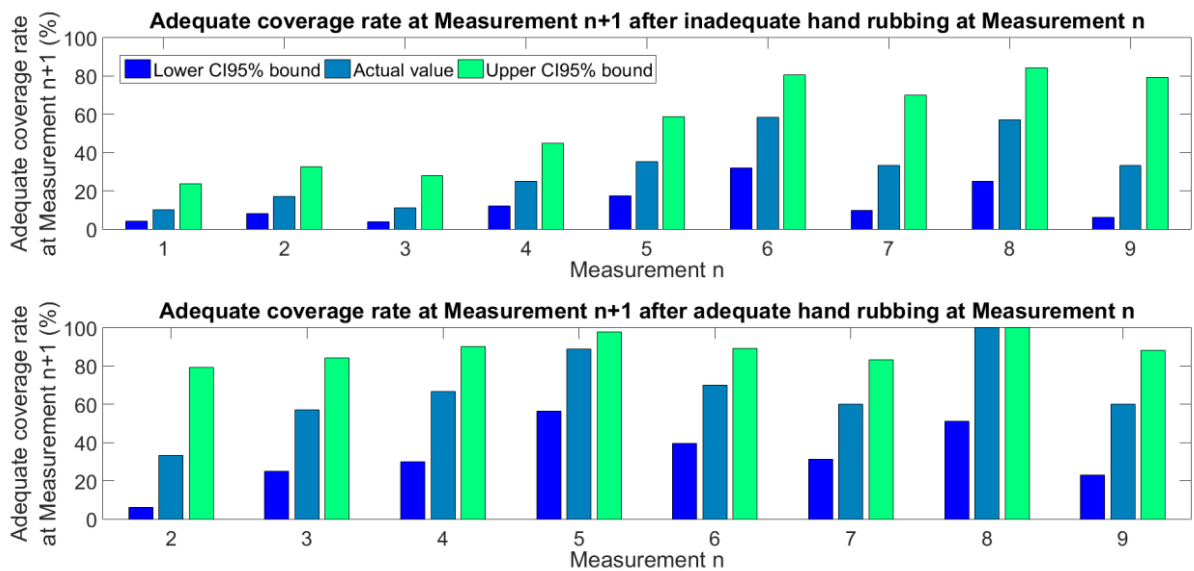

**Figure 5.** Statistics of consecutive measurements using 1.5 mL ABHR. The upper graph shows the adequate coverage rate at Measurement n + 1, occurring after inadequate coverage performed at Measurement n. The lower graph shows the adequate coverage rate at Measurement n + 1, occurring after adequate coverage performed at Measurement n.

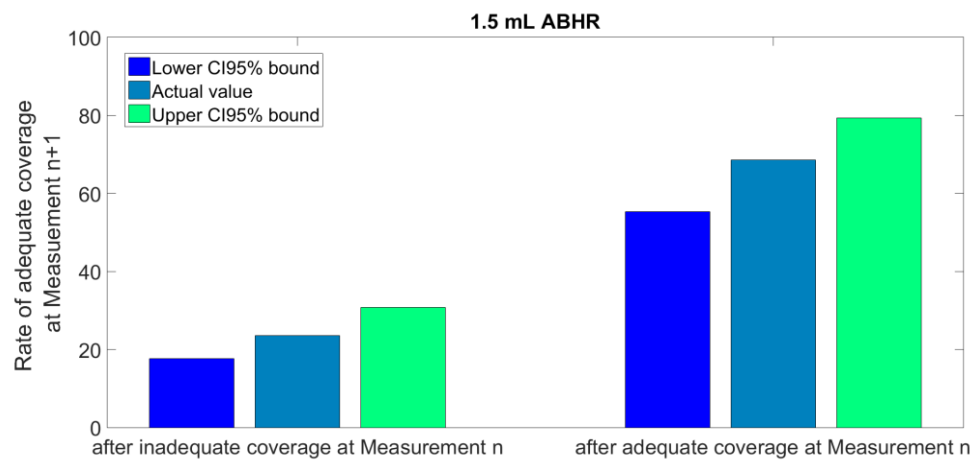

**Figure 6.** Overall statistics of consecutive measurements using 1.5 mL ABHR.

Participants were separated into two groups: we defined the criterion for being able to learn adequate hand rubbing technique as being able to produce perfect coverage at two consecutive measurements. According to Figure 4, 53% (CI 95%: 37–69%) of the participants could comply with this requirement. The two groups of participants shown in Figure 4, namely those who were able to produce perfect coverage at least two consecutive times and those who were not, have been compared with respect to the age, hand projection area, and the difference in coverage of the participant.

Table 2 presents the recorded numbers at the NICU regarding ABHR consumption and HAI ratio (for the period public data was made available). It has been recorded that the ABHR consumption stood out of the year of the study, rising to 157% of the 4-year average, while there was a drop down in the consecutive years. In the meanwhile, HAI numbers could not be correlated to the handrub use.

**Table 2.** Reported numbers of the NICU regarding handrub consumption and HAI. International recommendations denote >100 L/1000 patient days ABHR consumption.

| Year | ABHR Consumption (L) | Patient Days (PD) | L/1000 PD | Patients Tracked w/Microbiology Surveillance | Patients Affected by HAI | Number of Reported HAI | BSI within the HAI |
|---|---|---|---|---|---|---|---|
| **2015** | 591 | 3853.5 | 153.4 | 69 | 9 | 12 | 8 |
| **2016** | 530 | 2203.5 | 240.5 | 43 | 11 | 19 | 13 |
| **2017** | 620 | 4764 | 130.1 | 32 | 15 | 12 | 5 |
| **2018** | 698 | 6524.5 | 106.1 | 59 | 5 | 5 | 1 |

## 4. Discussion

Our study concludes that an electronic HH education and monitoring device enables approximately half of the HCWs to learn the adequate hand rubbing technique using only 1.5 mL ABHR solution, while 3 mL ABHR would be enough for every HCW. Previous studies in this matter formulated various opinions. It has been reported that 1 mL ABHR is hardly enough for any HCW to produce fine coverage [34]. Another study found that 3 mL ABHR is not enough for a perfect coverage [35], but their participants were volunteering visitors of an IPC conference who may not have adequate knowledge about HH technique, and did not perform repeated tests following the instant visual feedback received. Others studied the effect of hand rubbing via microbiological cultivation and bacterial colony counting [36]. They found that in case of large hands, up to 6 mL ABHR may be necessary for an efficient HH.

The findings of this study suggest that a volume of 1.5 mL ABHR is relative to be sufficient or not, depending not only on the size of the hands [37], but also the competency of the HCW, given the overwhelming effect of learning. HCWs having smaller hands are more likely to produce fine coverage, but the difference was non-significant. This result is in line with the findings of Zingg et al., who also found hand size to be a major but non-significant factor in the matter of the volume of ABHR necessary for fine hand coverage [35]. Recently, more precise algorithms have been able to assess the true size of human hands and relate performance to that [38].

The age of the HCWs shows a stronger correlation with the ability to learn adequate HH using 1.5 mL ABHR, but neither this relation is statistically significant, given the small number of participants. This observation is in line with the findings of Szilágyi et al., who established in a large-scale study that HCWs aged between 40 and 60 years are more likely to produce adequate coverage than those aged under 40 [10]. The results of one-way ANOVA tests are exhibited in Figure 7. The area of the hand did not show any reasonable difference between the two groups, and the average age was also insignificant (46 vs. 42 years, *p* = 0.248).

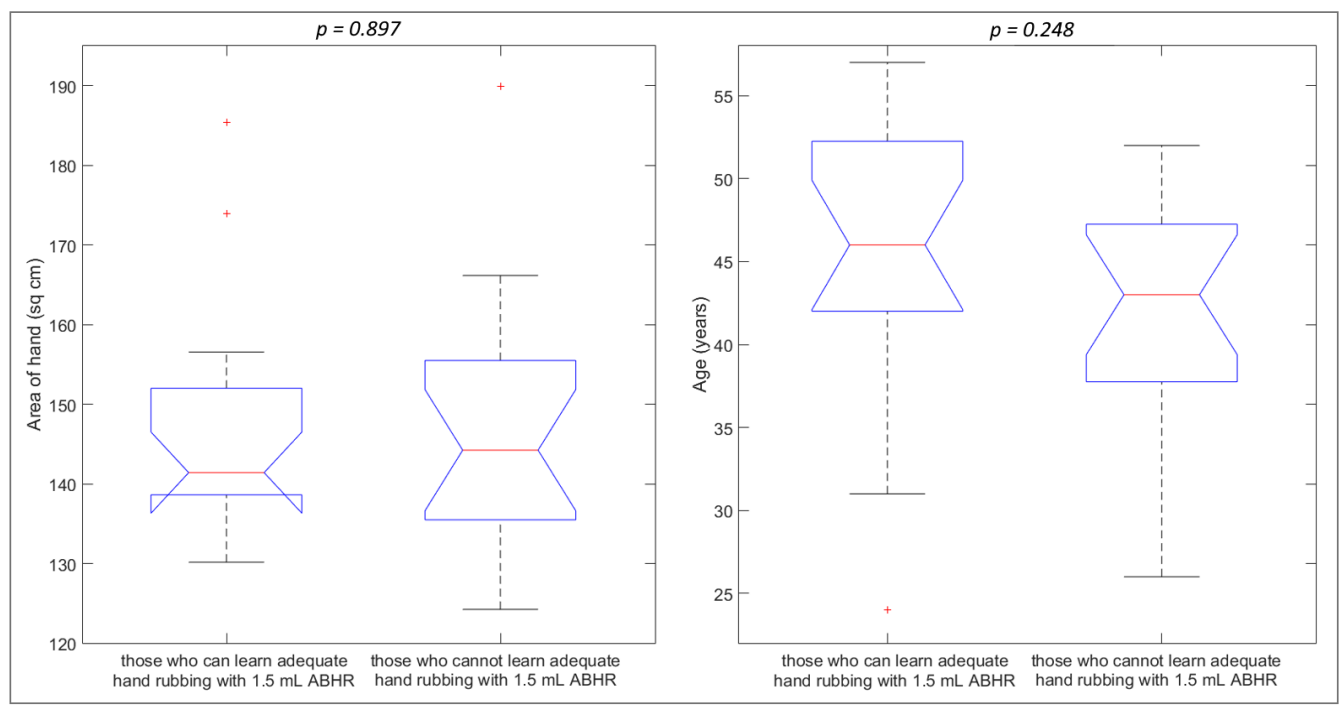

**Figure 7.** Relation between the ability to learn adequate hand rubbing technique and parameters like the area of hand and age of the HCW. Red crosses indicate outliers.

The results of this study suggest that there must be other factors (e.g., skin type, dryness of the hands) that influence the necessary ABHR quantity needed for a given HCW to provide adequate hand rubbing [37]. These factors may be difficult to identify, but certain electronic devices like the Semmelweis Scanner are able to track the volume of ABHR separately for each individual, thus contributing to the cost optimization of infection prevention actions.

Regarding the patient safety measures applicable to a NICU, the following assumptions are deduced:

- Low ABHR volume correlates with transmission risk, therefore monitoring ABHR application volume per HH event can be a proxy for quality outcome;
- With 1.5 mL ABHR only, the learning curve is slower, which imposes a patient risk at a NICUr
- The faster competence acquisition including 3 mL ABHR mean a saving on training time;
- Complete hand coverage is far from trivial even with 3 mL ABRH, therefore, regular skill training is required;
- It was observed that the quality of HH strongly correlated with the size of the hands, therefore larger hands are supposed to receive larger amounts of ABHR.

Optimized ABHR requirement thus means a financial benefit for the hospital, without endangering patient safety. The average price of clinical ABHR in Hungary costs before the pandemic EUR 2.48/L therefore the annual increase in handrub consumption during the study means a EUR 980 spending with a high possible upside regarding patient outcome.

Hungary maintains a public healthcare system, thus the Medical Center is deemed to optimize the use of its human and monetary resources, which are especially thin with the IPC. Serving as a teaching hospital as well, there is a continuous inflow of medical residents and junior practitioners, putting a high IP training load on the IPC staff. According to the latest publicly available National Nosocomial Surveillance System data, the average ABHR use was 10.7 L/1000 patient days (based on 22 institutions), which is merely half of the WHO recommended amount [39]. While the consumption went up significantly during the COVID-19, it is still significantly lagging behind international averages.

The focused training program would be able to identify and compensate skill deficits identified with individuals, e.g., after the return from a prolonged vacation or when moving into a new facility, or when changing departments. In the future we plan to extend it to visitors and patients as well, since the literature reports that there is an even bigger skill gaps those groups [26].

In the year of the training, the disinfectant consumption increased to the level deemed necessary to reach the appropriate HH, however, from the following years, we experienced a decrease again, which was not followed by an increase in the number of reported adverse events. Further research is needed to understand the relationship between the appropriate HH, hospital reported events (infection incidences) and disinfectant consumption.

## 5. Limitations and Future Work

This study is limited by the lack of a control group. Some participants only provided a few measurements, and it is possible that HCWs who were more highly motivated to improve their HH were more likely to perform numerous measurements. The study took place in a single department of a certain hospital. Only one ABHR was involved in the tests. The 1.5 mL units used by the ABHR dispenser represents a limitation in the accuracy of the estimated necessary ABHR volume for each participant. The hand projection area recording was prone to errors due to the drawing process performed by humans.

The duration and the inclusion of the study were both very limited, and the model has not yet been tested on data from other hospitals. Nevertheless, the digital records of the raw outcomes are available, which enables future retrospective evaluation and comparison of the data.

In the future, it would be important to increase the scope and range of the study, involving more healthcare workers for a longer period of time. It would be worthwhile to verify on the basis of the test data that those who have 2 (or more) consecutive O's (complete HH events), the number of errors decreases significantly afterwards, and it is thus a good indicator of learning success. This could be modelled, e.g., with a Poisson panel regression, where the effect of the event before/after two consecutive complete events could be examined on the frequency of occurrence of errors (rate), or with time-to-event analysis (e.g., multiple-event Cox regression—time elapsed until the next error, allowing repeated errors in the case of one respondent) to prove that this is a valid model. It would also be worth verifying whether there is a difference between 1 consecutive O or 2, 3, and 4, etc… complete events, as an indicator.

The examination of the cost-effectiveness of the Semmelweis Hand Hygiene System requires further research. We can assume that:

(a) It reduces the number of events—this means an increase in quality of life and a decrease in costs;
(b) Speeds up learning—this saves on personnel costs (but depends on staff turnover, etc.);
(c) Can increase disinfectant loss if too low—it is not known if it reduces wastage (it depends on the pattern of disinfectant loss).

## 6. Conclusions

Workplace overload and stress have become a critical issue with NICU staff on a global scale, leading to gaps in patient safety, in some cases [40]. The lack of time spent is often cited as one of the leading causes of non-compliance, with hand hygiene requirements. It is critical for the IPC staff to focus their limited human resources on trainings that take the greatest effect in the long term, reducing SSI ratios, and improving patient outcomes in general. To maximize the efficacy of staff training, the learning model of hand hygiene techniques was assessed, based on measurements in a Hungarian primary care facility's NICU. The study showed the effect of objective hand hygiene training in a clinical setup, and the learning model for individual healthcare workers was established. It has been shown that both in the case of 1.5 mL and 3 mL ABHR, the outcome of the HH, with respect to coverage, can be greatly improved with individual feedback, based on objective evaluation. Moreover, it has been shown that some people (depending on the skin type, hand size and other metrics) are able to learn proper disinfection. Conversely, some people are incapable of acquiring this skill, even after prolonged training sessions. A cutting-edge training model has to be individualized and can be made most efficient with direct feedback for the HCWs. The importance of training over the volume of handrub has been demonstrated, showing that, while 1.5 mL is typically not enough to cover hands, 3 mL can provide a complete coverage, with proper training. Secondary indicators, such as the rise in ABHR consumption in the ward, were confirmed, as a 157% increase in handrub consumption was linked to the training, which converges to the expected level ensuring adequate hand hygiene. However, the positive outcome with respect to the number of HAIs is yet to be demonstrated.

**Author Contributions:** Conceptualization, L.S. and T.H.; data curation, C.N.; formal analysis, L.S.; funding acquisition, T.H.; investigation, I.A.K.N. and Á.L.; methodology, I.A.K.N., C.N. and Á.L.; resources, C.N. and T.H.; software, L.S.; supervision, T.H.; validation, Á.L.; writing—original draft, L.S., T.H and Á.L.; writing—review & editing, C.N. and T.H. All authors have read and agreed to the published version of the manuscript.

**Funding:** The authors gratefully acknowledge the support of HandInScan Zrt., providing the Semmelweis Scanner and the technical background for the study. HARTMANN-RICO Hungária Kft. is recognized for their support with the training solution. This project has been partially supported by the National Research, Development, and Innovation Fund of Hungary, financed under the TKP2021-NKTA-36 funding scheme (Development and evaluation of innovative and digital health technologies—Evaluation of digital medical devices: efficacy, safety, and social utility).

**Institutional Review Board Statement:** Ethical review and approval were waived for this study due to the nature of it.

**Informed Consent Statement:** Not applicable.

**Data Availability Statement:** Date is available from the authors upon request.

**Acknowledgements**: All staff members of the NICU are thanked for their involvement. Special thanks to Zsombor Zrubka for his insightful comments and to Klara Haidegger for technical editing the manuscript.

**Conflicts of Interest:** Authors declare no conflict of interest. László Szilágyi, Ákos Lehotsky and Tamás Haidegger are co-founders of HandInScan Zrt.

*J. Clin. Med.* **2022**, *11*, 4276 14 of 1432. Bansaghi, S.; Haidegger, T. Towards Objective Hand Size Assessment and a Standardized Measurement Technique. In Proceedings of the 2020 IEEE 20th International Symposium on Computational Intelligence and Informatics (CINTI), Budapest, Hungary, 5–7 November 2020; pp. 139–144.
33. Macinga, D.R.; Shumaker, D.J.; Werner, H.P.; Edmonds, S.L.; Leslie, R.A.; Parker, A.E.; Arbogast, J.W. The relative influences of product volume, delivery format and alcohol concentration on dry-time and efficacy of alcohol-based hand rubs. *BMC Infect. Dis.* **2014**, *14*, 511.
34. Kampf, G.; Ruselack, S.; Eggerstedt, S.; Nowak, N.; Bashir, M. Less and less—Influence of volume on hand coverage and bactericidal efficacy in hand disinfection. *BMC Infect. Dis.* **2013**, *13*, 472.
35. Zingg, W.; Haidegger, T.; Pittet, D. Hand coverage by alcohol-based handrub varies: Volume and hand size matter. *Am. J. Infect. Control* **2016**, *44*, 1689–1691.
36. Bellissimo-Rodrigues, F.; Soule, H.; Gayet-Ageron, A.; Martin, Y.; Pittet, D. Should alcohol-based handrub use be customized to healthcare workers' hand size? *Infect. Control Hosp. Epidemiol.* **2016**, *37*, 219–221.
37. Voniatis, C.; Bánsághi, S.; Ferencz, A.; Haidegger, T. A large-scale investigation of alcohol-based handrub (ABHR) volume: Hand coverage correlations utilizing an innovative quantitative evaluation system. *Antimicrob. Resist. Infect. Control* **2021**, *10*, 49.
38. Voniatis, C.; Bánsághi, S.; Szerémy, P.; Veres, D.S.; Ferencz, A.; Jedlovszky-Hajdu, A.; Haidegger, T. Application differences of liquid and gel alcohol based handrubs formulations. In Proceedings of the Chemistry Physics and Biology of Colloids and Interfaces (CPBCI), Eger, Hungary, 6–10 June 2022; p. 94.
39. Ripabelli, G.; Tamburro, M.; Guerrizio, G.; Fanelli, I.; Agnusdei, C.P.; Sammarco, M.L. A single-arm study to evaluate skin tolerance, effectiveness and adherence to use of an alcohol-based hand rub solution among hospital nurses. *J. Infect. Prev.* **2019**, *20*, 224–230.
40. Rittenschober-Böhm, J.; Bibl, K.; Schneider, M.; Klasinc, R.; Szerémy, P.; Haidegger, T.; Ferenci, T.; Mayr, M.; Berger, A.; Assadian, O. The association between shift patterns and the quality of hand antisepsis in a neonatal intensive care unit: An observational study. *Int. J. Nurs. Stud.* **2020**, *112*, 103686.